\documentclass[conference]{IEEEtran}
\usepackage{epsfig,rotating,setspace,latexsym,amsmath,epsf,amssymb,bm}
\usepackage{cite,mathrsfs,authblk}

\IEEEoverridecommandlockouts

\title{On X-Channels with Feedback and Delayed CSI\thanks{The research of H. V. Poor and R. Tandon was supported in part by the Air Force Office of Scientific Research MURI Grant FA-$9550$-$09$-$1$-$0643$ and in part by the National Science Foundation Grant CNS-$09$-$05398$. The work of S. Shamai was supported by the Israel Science Foundation (ISF), and the Philipson Fund for Electrical Power. E-mail: \{rtandon, poor\}@princeton.edu, mohajer@eecs.berkeley.edu, sshlomo@ee.technion.ac.il.}
}

\author[1]{Ravi Tandon}
\author[2]{Soheil Mohajer}
\author[1]{H. Vincent Poor}
\author[3]{Shlomo Shamai}
\affil[1]{Dept. of EE, Princeton University, Princeton, NJ}
\affil[2]{Dept. of EECS, University of California at Berkeley, Berkeley, CA}
\affil[3]{Dept. of EE, Technion, Israel Institute of Technology, Haifa, Israel}

\newcommand{\DoF}{\mathbf{DoF}}

\newtheorem{Theo}{Theorem}
\newtheorem{Rem}{Remark}

\begin{document}
\maketitle
\begin{abstract}
The sum degrees of freedom ($\DoF$) of the two-user MIMO $X$-channel is characterized in the presence of output feedback
and delayed channel state information (CSI). The number of antennas at each transmitters is assumed to be $M$ and the
number of antennas at each of the receivers is assumed to be $N$. It is shown that
the sum $\DoF$ of the two-user MIMO $X$-channel is the {\emph{same}} as the sum $\DoF$ of a two-user MIMO broadcast channel
with $2M$ transmit antennas, and $N$ antennas at each receiver. Hence, for this symmetric antenna configuration, there is no performance loss
in the sum degrees of freedom due to the distributed nature of the transmitters. This result highlights the usefulness
of feedback and delayed CSI for the MIMO $X$-channel.

The $K$-user $X$-channel with single antenna at each transmitter and each receiver is also studied.
In this network, each transmitter has a message intended for each receiver. For this network, it is shown
that the sum $\DoF$ with {\emph{partial}} output feedback alone is at least $2K/(K+1)$. This lower bound
is strictly better than the best lower bound known for the case of delayed CSI assumption for \emph{all} values of $K$.
\end{abstract}

\section{Introduction}
In currently deployed wireless networks, multiple pairs of users wish to communicate with each other over a shared medium.
Due to the inherent broadcast nature of the wireless medium, interference is one of the main bottlenecks in efficient utilization of communication resources.
Several approaches to combat interference have been proposed in the literature such as treating interference
as noise, or decoding interference and subtracting it from the received signal. However, for multiple users, such approaches can be sub-optimal in general. Recently, more sophisticated schemes, such as interference alignment and (aligned) interference neutralization have been proposed for managing interference (see \cite{Jafar:Tutorial} for an excellent tutorial and the references therein). However, these techniques are usually based on availability of instantaneous (perfect) and global channel state information (CSI) at the transmitters. Such an assumption is perhaps not very realistic in practical systems, when dealing with fast fading links.

The pioneering work in \cite{MaddahAli-Tse:DCSI-BC} considers a model in which the perfect CSI assumption is relaxed to delayed CSI;
a setting in which CSI is available in a delayed manner at the transmitters. Interestingly, it is shown in \cite{MaddahAli-Tse:DCSI-BC}
that even delayed CSI can be helpful in increasing the degrees of freedom ($\DoF$) for broadcast multiple-input multiple output (MIMO) networks, even if the
channel changes independently over time. Several interesting extensions of \cite{MaddahAli-Tse:DCSI-BC} have been considered recently; which include the two-user MIMO broadcast channel (BC) \cite{VV:DCSI-BC}, the three user MIMO-BC \cite{AbdoliISIT3user,VV:DCSI-BC} and the two-user MIMO interference channel (IC) \cite{VV:DCSI-IC}.

 A very relevant question is that whether channel output feedback (FB) can be helpful with delayed CSI or not. For the case of the MIMO-BC, this question is answered in a negative way in \cite{VV:DCSI-BC}: i.e., having output feedback, in addition to delayed CSI does not increase the $\DoF$ region of the MIMO-BC, even though it enlarges the capacity region. However, FB if available in addition to delayed CSI can increase the $\DoF$ for the MIMO interference channel (MIMO-IC). This is shown explicitly in \cite{TMPS:DCSIFB-IC}, where the $\DoF$ region of the two-user MIMO-IC is completely characterized
in the presence of FB and delayed CSI (also see the parallel work in \cite{VV:DCSIFB-IC}, which reports similar results).

The study of the impact of delayed CSI on the $\DoF$ of $X$-channels was initiated in \cite{MJS:DCSI-X}. It is shown that for the two-user
$X$-channel, a sum $\DoF$ of $8/7$ is achievable with delayed CSI. The impact of FB on the sum $\DoF$ for the two-user X-channel and the
three-user IC is also explored in \cite{MJS:DCSI-X}. It is shown that a sum $\DoF$ of $6/5$ is achievable for the
three-user IC with FB alone. Moreover, the optimal sum $\DoF$ of the two-user $X$-channel with FB
is shown to be $4/3$. It is worth noting that for the single-antenna two-user $X$-channel, FB alone is sufficient to
achieve the outer bound of $4/3$, which  holds also for the stronger setting of FB and delayed CSI.

The focus of this paper is on MIMO $X$-channels with output feedback and delayed CSI. The sum $\DoF$ of the MIMO
$X$-channel is characterized for the symmetric antenna configuration, with $M$ antennas at each transmitter and $N$ antennas at each receiver.
It is shown that the sum $\DoF$ of the MIMO $X$-channel equals the sum $\DoF$ of a MIMO-BC with $2M$ transmit antennas
and $N$ antennas at each of the receivers. This result highlights the fact that in the presence of output feedback and delayed CSI, there is no $\DoF$ loss due to the distributed nature of the $M$-antenna transmitters.

We also focus on the setting of the $K$-user $X$-channel with a single antenna at each terminal. For this model under the assumption of global output feedback and delayed CSI, the sum $\DoF$ is also shown to be the same as that of a $K$-receiver MISO-BC \cite{MaddahAli-Tse:DCSI-BC} with $K$ transmit antennas, i.e., $K/(1+\frac{1}{2}+\ldots+\frac{1}{K})$.
The assumption of global feedback is then relaxed to partial (local) feedback, in which receiver $j$ sends feedback only to transmitter $j$, for $j=1,\ldots,K$. For this model, it is shown that the sum $\DoF$ is lower bounded by $2K/(K+1)$. The interest of this lower bound is that it is strictly larger than the best known lower bound for the delayed CSI setting \cite{Khandani:DCSIT}.
Finally, for the $K$-user IC with a single antenna at each terminal and global feedback and delayed CSI, the sum $\DoF$ is shown to be lower bounded by $K/(2+\frac{1}{2}+\ldots+\frac{1}{K})$. Interestingly, this shows that for large values of $K$, the behavior of the sum $\DoF$ of the $K$-user BC, $K$-user $X$-channel and the $K$-user IC is similar in the presence of global FB and delayed CSI.

\section{MIMO X-channel with FB and Delayed CSI}
\begin{figure}[t]
 \centerline{\epsfig{figure=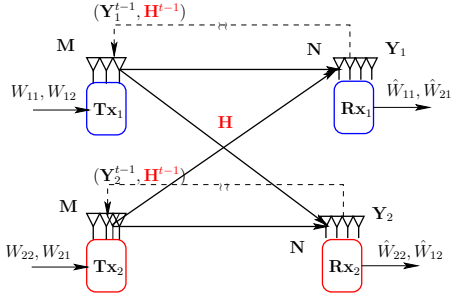,width=6.0cm}}
  \caption{The MIMO X-channel with output feedback and delayed CSI.}\label{Figmodel}
  \vspace{-0.7cm}
\end{figure}

We consider the two-user $(M,M,N,N)$-MIMO X-channel with fast fading under the assumptions of
(A-I) noiseless channel output feedback from receiver $n$ to transmitter $n$, for $n=1,2$
and (A-II) the availability of delayed CSI at the transmitters (see Figure \ref{Figmodel}).  We denote the transmitters by
$\mathbf{Tx}_{1}$ and $\mathbf{Tx}_{2}$ and the receivers by $\mathbf{Rx}_{1}$ and $\mathbf{Rx}_{2}$.
The channel outputs at the receivers are given as follows:
\begin{align}
Y_{1}(t)&= \mathbf{H}_{11}(t)X_{1}(t)+ \mathbf{H}_{12}(t)X_{2}(t)+ Z_{1}(t)\nonumber\\
Y_{2}(t)&= \mathbf{H}_{21}(t)X_{1}(t)+ \mathbf{H}_{22}(t)X_{2}(t)+ Z_{2}(t)\nonumber,
\end{align}
where $X_{n}(t)$ is the signal transmitted by $n$th transmitter $\mathbf{Tx}_{n}$; $\mathbf{H}_{ij}(t)\in \mathbb{C}^{N\times M}$
denotes the channel matrix between the $i$th receiver and $j$th transmitter; and $Z_{n}(t)\sim \mathcal{CN}(0,I_{N})$, for $n=1,2$,
is the additive noise at receiver $n$. The power constraints are $\mathbb{E}||X_{n}(t)||^{2}\leq P$, for $\forall$  $n,t$.

For the $X$-channel, there are four independent messages, one from each transmitter to each receiver. In particular,
we denote by $W_{i,j}$ the message from transmitter $i$ to receiver $j$. We denote by $\mathbf{H}(t)= \{\mathbf{H}_{11}(t),\mathbf{H}_{12}(t),\mathbf{H}_{21}(t),\mathbf{H}_{22}(t)\}$ the collection of all channel matrices at time $t$. Furthermore, $\mathbf{H}^{t-1}=\{\mathbf{H}(1),\mathbf{H}(2),\ldots,\mathbf{H}(t-1)\}$ denotes the set of all channel matrices up till time $(t-1)$. Similarly, we denote by $Y_{n}^{t-1}=\{Y_{n}(1),\ldots,Y_{n}(t-1)\}$ the set of all channel outputs at receiver $n$ up till time $(t-1)$.
A coding scheme with block length $T$ for the MIMO X-channel with feedback and delayed CSI consists of a sequence of encoding functions
\begin{align}
X_{1}(t)&=f^{T}_{1,t}\left(W_{11},W_{12},\mathbf{H}^{t-1},Y_{1}^{t-1}\right)\nonumber\\
X_{2}(t)&=f^{T}_{2,t}\left(W_{22},W_{21},\mathbf{H}^{t-1},Y_{2}^{t-1}\right)\nonumber,
\end{align}
defined for $t=1,\ldots,T$, and four decoding functions
\begin{align}
\hat{W}_{11}= g^{T}_{11}(Y_{1}^{n},\mathbf{H}^{n}),\quad \hat{W}_{21}= g^{T}_{21}(Y_{1}^{n},\mathbf{H}^{n}),\nonumber\\
\hat{W}_{22}= g^{T}_{22}(Y_{2}^{n},\mathbf{H}^{n}),\quad \hat{W}_{12}= g^{T}_{12}(Y_{2}^{n},\mathbf{H}^{n}).\nonumber
\end{align}
A rate quadruple $(R_{11}(P),R_{12}(P), R_{22}(P), R_{21}(P))$ is achievable if there exists a
sequence of coding schemes such that $\mathbb{P}(W_{ij}\neq \hat{W}_{ij})\rightarrow 0$
as $T\rightarrow \infty$ for all $(i,j)$.
The capacity region $\mathcal{C}(P)$ is defined as the set of all achievable rate pairs $(R_{11}(P),R_{12}(P), R_{22}(P), R_{21}(P))$.
We define the $\DoF$ region as follows:
\begin{align}
&\mathbf{D}=\Big\{(d_{11},d_{12},d_{22},d_{21})\bigg| d_{i,j}\geq 0, \mbox{ and } \nonumber\\
&\exists (R_{11}(P),R_{12}(P),R_{22}(P),R_{21}(P))\in \mathcal{C}(P) \nonumber\\
&\mbox{s.t. } d_{i,j}=\lim_{P\rightarrow \infty}\frac{R_{i,j}(P)}{\log_{2}(P)}, (i,j)=(1,1), (1,2), (2,2),(2,1) \Big\}.\nonumber
\end{align}

We denote the total (sum) degrees of freedom as $\DoF_{\mathrm{sum}}$, defined as
\begin{align}
\DoF_{\mathrm{sum}}(M,N)&= \max_{(d_{11},d_{12},d_{22},d_{21})\in \mathbf{D}} d_{11}+d_{12}+d_{22}+d_{21}.\nonumber
\end{align}
We first state an outer bound for the $\DoF$ region with feedback and delayed CSI:
\begin{align}
&\frac{d_{11}+d_{21}}{\min(2M,2N)} + \frac{d_{22}+d_{12}}{\min(2M,N)}\leq 1\label{OB1}\\
&\frac{d_{11}+d_{21}}{\min(2M,N)} + \frac{d_{22}+d_{12}}{\min(2M,2N)}\leq 1.\label{OB2}
\end{align}
This bound follows from \cite{VV:DCSI-BC} by letting the transmitters cooperate and subsequently
using the bound for the MIMO broadcast channel with feedback and delayed CSI.

We present our main result in the following theorem.
\begin{Theo}\label{Theorem1}
The sum $\DoF$ of the $(M,M,N,N)$-MIMO X-channel with feedback and delayed CSI is given as follows:
\begin{align}
\DoF_{\mathrm{sum}}(M,N)=
\left\{
  \begin{array}{ll}
    2M, &2M\leq N; \\
    \frac{4MN}{2M+N}, & N\leq 2M\leq 2N; \\
    \frac{4N}{3}, & 2N\leq 2M.
  \end{array}
\right.
\end{align}
\end{Theo}
The converse follows immediately from the MIMO broadcast channel bounds in (\ref{OB1}) and (\ref{OB2}).
We complete the proof for Theorem \ref{Theorem1} in the next section by presenting coding schemes with
feedback and delayed CSI. We note that for $M=N=1$, Theorem \ref{Theorem1} recovers a result of \cite{MJS:DCSI-X}, in which
the sum $\DoF$ of the single antenna $X$ channel was shown to be $4/3$.

\subsection{Coding Schemes}
\subsubsection{Coding scheme for $2M\leq N$}
For $2M\leq N$, we have $\DoF_{\mathrm{sum}}=2M$. We shall only outline the coding scheme since it is straightforward.
To this end, we will show the achievability of the following quadruple:
\begin{align}
(d_{11},d_{12},d_{22},d_{21})=(M,0,0,M),
\end{align}
which implies that there are $M$ symbols from transmitter $1$ and $M$ information symbols from transmitter $2$, both intended for receiver $1$.
Hence, coding for this system is equivalent to coding for a MIMO multiple access channel, for which the achievability follows from standard results.

\subsubsection{Coding scheme for $N\leq 2M\leq 2N$}
For this case, we have $\DoF_{\mathrm{sum}}=\frac{4MN}{2M+N}$. We present an encoding scheme that achieves the following
quadruple $(d_{11},d_{12},d_{22},d_{21})$:
\begin{align}
\left(\frac{MN}{2M+N},\frac{MN}{2M+N},\frac{MN}{2M+N},\frac{MN}{2M+N}\right),
\end{align}
that is, there are $MN$ information symbols at each transmitter for each receiver to be sent over $(2M+N)$ channel uses.
Let us denote
\begin{align}
\mathbf{u}_{11}=[u^{1}_{11},\ldots, u^{MN}_{11}], \quad \mathbf{u}_{21}=[u^{1}_{21},\ldots, u^{MN}_{21}],
\end{align}
as the symbols intended for receiver $1$, and
\begin{align}
\mathbf{v}_{22}=[v^{1}_{22},\ldots, v^{MN}_{22}],\quad \mathbf{v}_{12}=[v^{1}_{12},\ldots, v^{MN}_{12}]
\end{align}
as the symbols intended for receiver $2$.
Note that the symbols $\left(\mathbf{u}_{11},\mathbf{v}_{12}\right)$ are present at transmitter $1$,
and the symbols $\left(\mathbf{u}_{21},\mathbf{v}_{22}\right)$ are present at transmitter $2$, i.e., the origin of information
 symbols is distributed in contrast to the MIMO broadcast channel.
The scheme operates over three phases described as below:

\textbf{\emph{Phase $1$}}: This phase uses $N$ channel uses. In every channel use, transmitter $1$ sends fresh information symbols
for receiver $1$, and transmitter $2$ sends fresh information symbols intended for receiver $1$. Note that, our choice of the duration
for this phase guarantees that all $2MN$ information symbols in $\mathbf{u}_{11}$ and $\mathbf{u}_{21}$ are transmitted exactly once and at one antenna.
At the end of phase $1$, receiver $1$ has $N^{2}$ linearly independent equations in $2MN$ variables. Whereas, receiver $2$ has $N^{2}$ linearly
independent equations in the same $2MN$ $\mathbf{u}$-variables.

At the end of phase $1$, receiver $1$ requires $2MN- N^{2}$ additional equations in $\mathbf{u}$-variables for successful decoding of $2MN$ information symbols.
Note that upon receiving feedback from receiver $2$, transmitter $2$ has access to $N^{2}$ additional equations in the $\mathbf{u}$-variables. Since $2M\leq 2N$,
we have $2MN- N^{2}\leq N^{2}$, i.e., transmitter $2$ has enough information, which if somehow can be supplied to receiver $1$ will guarantee successful decoding of the $\mathbf{u}$-symbols. Let us denote these $(2MN-N^{2})$ symbols by $\widetilde{\mathbf{u}}$. More importantly, upon receiving feedback from receiver $1$, transmitter $1$
can subtract out the contribution from $\mathbf{u}_{11}$, and decode $\mathbf{u}_{21}$ (this is possible since $\mathbf{u}_{21}$ has $M$ symbols, the feedback vector is of length $N$ and we have $M\leq N$). Subsequently, having the CSI of the first block, transmitter $1$ can reconstruct the side-information $\widetilde{\mathbf{u}}$ available at transmitter $2$. To summarize, feedback and delayed CSI serve a dual purpose for this setting: not only does it provide side-information at transmitter $2$ (for future use), it also lets transmitter $1$ reconstruct the same side-information.

 \textbf{\emph{Phase $2$}}: This phase uses $N$ channel uses. In every channel use, transmitter $1$ sends fresh information symbols
for receiver $2$, and transmitter $2$ sends fresh information symbols intended for receiver $2$. At the end of phase $2$, receiver $2$ has $N^{2}$
linearly independent equations in $2MN$ variables $\mathbf{v}_{22}$ and $\mathbf{v}_{12}$. Whereas, receiver $1$ has $N^{2}$ linearly
independent equations in the same $2MN$ $\mathbf{v}$-variables. Similar to phase $1$, at the end of phase $2$, receiver $2$ requires $2MN- N^{2}$ additional equations in $\mathbf{v}$-variables for successful decoding of $2MN$ information symbols. Furthermore, upon receiving feedback from receiver $1$, transmitter $1$ has access to $N^{2}$ additional equations in $\mathbf{v}$-variables. Since $2M\leq 2N$, we have $2MN- N^{2}\leq N^{2}$, i.e., in this case transmitter $1$ has enough information, which if somehow can be supplied to receiver $2$ will guarantee successful decoding of $\mathbf{v}$-symbols. Let us denote these $(2MN-N^{2})$ by $\widetilde{\mathbf{v}}$. Similar to the end of phase $1$, transmitter $2$ can also reconstruct the side-information $\widetilde{\mathbf{v}}$-symbols.
At the end of this phase, {\emph{both}} transmitters $1$ and $2$ have access to the side-information symbols $(\widetilde{\mathbf{u}}, \widetilde{\mathbf{v}})$.
This is the key step behind the achievability proof, i.e., the {\emph{common}} availability of side-information symbols before phase $3$.

 \textbf{\emph{Phase $3$:}} This phase operates over $(2M-N)$ channel uses. The goal is to send $\widetilde{\mathbf{u}}$ to receiver $1$ and $\widetilde{\mathbf{v}}$ to receiver $2$. Note that from phase $1$, receiver $2$ has access to $\widetilde{\mathbf{u}}$, and from phase $2$, receiver $1$ has access to $\widetilde{\mathbf{v}}$.  Recall that each $\widetilde{\mathbf{u}}$ and $\widetilde{\mathbf{v}}$ are of length $2MN-N^{2}$.
Let us denote
\begin{align}
\widetilde{\mathbf{u}}=[\tilde{u}_{1},\ldots, \tilde{u}_{2MN-N^{2}}],\quad  \widetilde{\mathbf{v}}&=[\tilde{v}_{1},\ldots, \tilde{v}_{2MN-N^{2}}].
\end{align}
Using these, both transmitters can compute
\begin{align}
\widetilde{\mathbf{uv}}&=[\tilde{u}_{1}+\tilde{v}_{1},\ldots, \tilde{u}_{2MN-N^{2}}+\tilde{v}_{2MN-N^{2}}],
\end{align}
which is the element-wise summation of the $\widetilde{\mathbf{u}}$ and $\widetilde{\mathbf{v}}$ sequences.
The transmitters send each of these symbols exactly once on an antenna. In particular, we have a total of $2M$ transmit antennas
and $(2M-N)$ channel uses, i.e., this scheme is feasible as long as
\begin{align}
2MN-N^{2}&\leq 2M(2M-N)
\end{align}
which is true since $N\leq 2M$. At the end of phase $3$, receiver $1$ gets $(2M-N)N$ equations in $2(2MN-N^{2})$ variables.
However, receiver $1$ already knows half of these variables, namely $\tilde{v}$ variables from phase $2$, and hence it is left with
$2MN-N^{2}$ equations in $2MN-N^{2}$ $\tilde{\mathbf{u}}$-variables. Using these and the information from phase $1$ (i.e., $N^{2}$ equations of phase $1$),
receiver $1$ can decode all $2MN$ symbols. Similarly, decoder $2$ can also decode a total of $2MN$ information symbols.

To illustrate this scheme by an example, consider the case when $N=3$, $M=2$,  and
 $\DoF_{\mathrm{sum}}=4MN/(2M+N)= 24/7$.
Phase $1$ operates over $3$ channel uses and at its end, receiver $1$ has $9$ equations in $12$ $\mathbf{u}$-variables ($6$ originating from transmitter $1$
and $6$ from transmitter $2$).
Similarly, at the end of phase $2$, receiver $2$ has $9$ equations in $12$ $\mathbf{v}$-variables ($6$ originating from transmitter $1$
and $6$ from transmitter $2$).
For this example, there are $(2MN-N^{2})= 3$ side-information symbols intended for receiver $1$ (let us denote these additional symbols by $\tilde{u}_{1},\tilde{u}_{2},\tilde{u}_{3}$) and three side information symbols for receiver $2$ (denoted these symbols by $\tilde{v}_{1},\tilde{v}_{2},\tilde{v}_{3}$).
In phase $3$, which is of duration $(2M-N)= 1$, transmitters $1$ and $2$ send
\begin{align}
X_{1}(3)=
\left[
  \begin{array}{c}
    \tilde{u}_{1}+\tilde{v}_{1} \\
    \tilde{u}_{2}+\tilde{v}_{2} \\
  \end{array}
\right], \quad X_{2}(3)=
\left[
  \begin{array}{c}
    \tilde{u}_{3}+\tilde{v}_{3} \\
     \phi  \\
  \end{array}
\right],
\end{align}
where $\phi$ denotes a constant symbol. Also, receiver $1$ has $\tilde{v}_{1},\tilde{v}_{2}, \tilde{v}_{3}$
from phase $2$ and similarly, receiver $2$ has access to $\tilde{u}_{1},\tilde{u}_{2}, \tilde{u}_{3}$
from phase $1$. Therefore, phase $3$ (which is of duration $2M-N=1$) essentially provides receiver $1$ with $(2M-N)N=3$ equations in $\tilde{u}_{1},\tilde{u}_{2},\tilde{u}_{3}$
symbols and similarly, receiver $2$ gets $(\tilde{v}_{1},\tilde{v}_{2},\tilde{v}_{3})$. Using information from phases $1$ and $3$, receiver $1$ can decode $12$ information symbols. Similarly, using information from phases $2$ and $3$, receiver $2$ is able to decode $12$ information symbols.

\begin{Rem}
We now give the intuition as to why the total $\DoF$ for the MIMO $X$-channel turns out to be the same
as that for the MIMO broadcast channel. This is illuminated in phase $3$, which requires complimentary
broadcasting of side-information symbols. In particular, we need to transmit $2MN-N^{2}$ symbols to receiver $1$ and $2MN-N^{2}$
symbols to receiver $2$. However, to attain this goal, we have a total of $(2M-N)$ channel uses allotted for phase $3$ and \emph{distributed} transmitters
equipped with $M$ antennas each.  The feasibility of this scheme is crucially dependent on the omniscience of these side-information symbols at both transmitters. As we have shown, feedback and delayed CSI guarantee the common availability of these side-information symbols at both the
transmitters, effectively creating a $2M$-antenna MIMO broadcast channel for phase $3$.
\end{Rem}

\subsubsection{Coding scheme for  $2N\leq 2M$}
For this case, we have $\DoF_{\mathrm{sum}}=4N/3$.
The scheme for this case is a simple variation as in the previous section and only output feedback suffices.
In the first channel use, transmitters $1$ and $2$ send fresh information symbols intended for receiver $1$ on
$N$ antennas (which is possible since $N\leq M$). In the second channel use, transmitters $1$ and $2$ send fresh information symbols intended for receiver $2$ on
$N$ antennas. Each receiver has $N$ equations in $2N$ variables, and each receiver requires $N$ more equations for successful decoding.
In the third channel use, transmitter $1$ uses feedback from second channel use (which is side information for receiver $1$) and transmitter $2$ uses feedback from the first channel use (which is side information for receiver $1$). It is clear that at the end of three channel uses, each receiver can decode $2N$ symbols.

\section{$K$-user X-channel with Output Feedback}
In this section, we focus on the $K$-user $X$-channel.
In this model, we assume that each transmitter and receiver is equipped with a single antenna.
We study two models with different assumptions on the availability of feedback signals:
a) \emph{Global feedback:} channel output feedback is present from all $K$ receivers to all $K$ transmitters, and
b) \emph{Partial feedback:} transmitter $k$ receives feedback only from receiver $k$.

\begin{Theo}\label{Theorem2}
The sum $\DoF$ of the $K$-user $X$-channel with global feedback is given
as follows:
\begin{align}
\DoF_{\mathrm{sum}}^{K,\mathrm{global}}&= \frac{K}{1+\frac{1}{2}+\ldots+\frac{1}{K}}.
\end{align}
\end{Theo}
Note that the sum $\DoF$ with global feedback is the same as the sum $\DoF$ for a multiple-input single output (MISO) broadcast channel with
$K$ transmit antennas and $K$ single antenna receivers.
The proof of Theorem \ref{Theorem2} is rather straightforward and is immediate from \cite{MaddahAli-Tse:DCSI-BC}. To note this,
we can proceed by using a coding scheme consisting of several phases.
The first phase is comprised of $K$ channel uses. During the first phase, in each channel use, all transmitters send information
for a fixed receiver. At the end of phase $1$, upon receiving \emph{global} feedback, each transmitter can decode all information symbols,
thus creating a virtual MISO broadcast channel. The coding for the subsequent phases follows as in \cite{MaddahAli-Tse:DCSI-BC}.

In the following theorem, we state a lower bound on the sum $\DoF$ for the $K$-user $X$-channel with partial feedback.
\begin{Theo}\label{Theorem3}
The sum $\DoF$ of the single-antenna $K$-user $X$-channel with partial feedback
is lower bounded as follows:
\begin{align}
\DoF_{\mathrm{sum}}^{K,\mathrm{partial}}&\geq \frac{2K}{K+1}.
\end{align}
\end{Theo}
We note that this bound is tight for $K=2$, for which we achieve the MISO broadcast channel bound of $4/3$ \cite{MJS:DCSI-X}.
We also note here that unlike the case for global feedback, the lower bound with
partial feedback {\emph{does not}} scale with $K$, the number of users.
Nevertheless, the lower bound stated in Theorem \ref{Theorem3} is strictly better than the
best known lower bound for the case with delayed CSI alone \cite{Khandani:DCSIT} for all values of $K$.
However, without a matching converse, we cannot claim the optimality of this lower bound.
\begin{figure*}[t]
 \centerline{\epsfig{figure=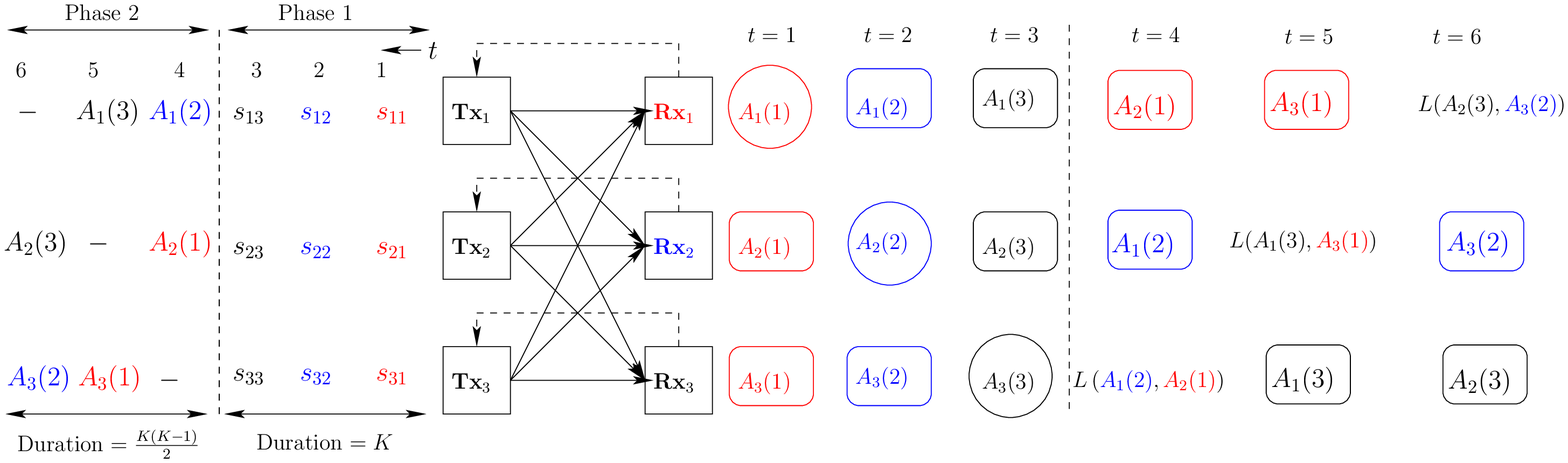,width=13.0cm, height=3.6cm}}
  \vspace{-0.2cm}
  \caption{Coding for the $3$-user X-channel with partial feedback.}\label{Figcodingexample1}
  \vspace{-0.6cm}
\end{figure*}

\subsection{Coding for the $3$-user X-channel}
Before presenting the proof for Theorem \ref{Theorem3}, we illustrate the coding scheme for the $K=3$ user X-channel.
For this case, we will show the achievability of $9/6$, i.e., we show that a total of $9$ information
symbols can be transmitted in $6$ channel uses. We denote an information symbol as $s_{ij}$ if it originates at transmitter $i$
and is intended for receiver $j$, $i,j=1,2,3$. In particular, for this example, we have the following notation:
\begin{itemize}
  \item $s_{11},s_{21},s_{31}$: symbols intended for receiver $1$.
  \item $s_{12},s_{22},s_{32}$: symbols intended for receiver $2$.
  \item $s_{13},s_{23},s_{33}$: symbols intended for receiver $3$.
\end{itemize}
Transmission occurs over two phases. Phase $1$ is of duration $3$, and phase $2$ is duration $3$ (see Figure \ref{Figcodingexample1}).

 \textbf{\emph{Phase $1$:}} During this phase, at time $t$, transmitters $1$, $2$ and $3$, send $s_{1t},s_{2t}$ and $s_{3t}$ respectively, for $t=1,2,3$.
 Note that for recovery of the three symbols $s_{11},s_{21},s_{31}$ at receiver $1$, two more linearly independent equations are required. These correspond to symbols $(A_{2}(1),A_{3}(1))$ which need to be delivered to receiver $1$. Similarly, the symbol pair $(A_{1}(2),A_{3}(2))$ needs to be delivered to receiver $2$, and the symbol pair $(A_{1}(3),A_{2}(3))$ to receiver $3$.

   \textbf{\emph{Phase $2$:}} In this phase we will show that it is possible to deliver the two complementary symbols to each of the respective receiver in three channel uses. Due to \emph{partial} feedback from phase $1$, transmitter $j$ has access to $A_{j}(1),A_{j}(2)$ and $A_{j}(3)$, for $j=1,2,3$.
      The coding in this phase works as follows:  at $t=4$, transmitter $1$ sends $A_{1}(2)$ and transmitter $2$ sends $A_{2}(1)$, whereas transmitter $3$ remains silent. This enables receiver $1$ to obtain $A_{1}(2)$ and receiver $2$ to obtain $A_{1}(2)$. At $t=5$, transmitter $1$ sends $A_{1}(3)$, transmitter $3$ sends $A_{3}(1)$ and transmitter $2$ remains silent. This enables receiver $1$ to obtain $A_{3}(1)$ and receiver $3$ to obtain $A_{1}(3)$. Finally, at $t=6$, transmitter $2$ sends $A_{2}(3)$, transmitter $3$ sends $A_{3}(2)$ and transmitter $1$ remains silent. Consequently,
      receiver $2$ gets $A_{3}(2)$ and receiver $3$ gets $A_{2}(3)$. Hence, at the end of this phase, each receiver has $3$ linearly independent equations in $3$ information symbols and the decoding is successful.

\subsection{Coding for the $K$-user X-channel}
To show the achievability of $2K/(K+1)$, we will show that it is possible to transmit
$K^{2}$ symbols in $K+ \frac{K(K-1)}{2}$ channel uses. As in case for $K=3$, there are two phases. Phase $1$ is of duration $K$, in which, at time $t$, each transmitter sends an information symbol
intended for receiver $t$, for $t=1,\ldots,K$. Hence, a total of $K^{2}$ symbols are transmitted over this phase, with a total of
$K$ symbols intended for each receiver. At the end of this phase, each receiver requires $(K-1)$ additional equations
for decoding the $K$ information symbols, i.e., there are a total of $K(K-1)$ additional symbols to be delivered.
Mimicking the scheme for $K=3$, we create pairs of these $K(K-1)$ symbols and reliably transmit these in $K(K-1)/2$ channel uses. Hence,
phase $2$ is of duration $K(K-1)/2$.
Therefore, this scheme can achieve the following sum $\DoF$:
\begin{align}
\DoF_{\mathrm{sum}}^{K,\mathrm{partial}}&\geq \frac{K^{2}}{\big[K+ \frac{K(K-1)}{2}\big]}= \frac{2K}{K+1}.
\end{align}
We note here that the proposed scheme {\emph{only}} requires channel output feedback from receiver $j$ to transmitter $j$, and no CSI (not even delayed)
is required at any of the transmitters.

\section{$K$-user IC: Feedback and Delayed CSI}
In this section, we focus on the $K$-user IC.
Given the scaling behavior for the $X$-channel in Theorem \ref{Theorem2},
a natural question arises: does the sum $\DoF$ for the $K$-user IC scale with $K$ in the presence of global feedback
and delayed CSI? We answer this question in the affirmative in the following theorem.
\begin{Theo}\label{Theorem4}
The sum $\DoF$ of the $K$-user IC with global feedback and delayed CSI is lower bounded as follows:
\begin{align}
\DoF^{\mathrm{IC}}(K)&\geq \frac{K}{\big[2+ \frac{1}{2}+\frac{1}{3}+\ldots+\frac{1}{K}\big]}
\end{align}
\end{Theo}
To show the achievability, we operate over two phases. In the first phase, all
transmitters send information symbols simultaneously. The output at receiver $j$ is a combination of the symbol from transmitter $j$
and an interference component, $I_{j}$, which is combination of the other $(K-1)$ symbols. Via global feedback and delayed CSI, all the $K$ interference
components $\{I_{1},\ldots,I_{K}\}$ can be recovered at each of the transmitters, thus creating a virtual MISO-BC.
In the next phase, we use the scheme of \cite{MaddahAli-Tse:DCSI-BC} to send the component $I_{j}$ to receiver $j$. Hence, the rate of this scheme is given as $K/(1 + K/\DoF^{BC}(K))$. As a consequence of Theorems \ref{Theorem2} and \ref{Theorem4}, the behavior of the MISO-BC, the $K$-user X-channel and the $K$-user IC are the same for large values of $K$.

\vspace{-0.05in}
\section{Conclusions}
The usefulness of feedback when available in addition to delayed CSI is illustrated by showing that the sum $\DoF$ of the symmetric MIMO X-channel is the same as the sum $\DoF$ of the MIMO-BC. A similar result is also shown for the $K$-user single-antenna $X$-channel. The result of Theorem \ref{Theorem3} also shows that partial output feedback yields a larger sum $\DoF$ when compared to the setting of delayed CSI. Moreover, it is shown that the scaling behavior of the sum $\DoF$ for the $K$-user IC in the presence of global feedback and delayed CSI is the same as that of the $K$-user MISO-BC.
\vspace{-0.05in}

\bibliographystyle{unsrt}
\bibliography{refravi}
\end{document}